# Enhanced Image Analysis Using Cached Mobile Robots.

Kabeer Mohammed
Dr.Bhaskara Reddy Ph.D(C.S)
Associate Professor
Department of Computer Science and Technology.
Sri Krishnadevaraya University

**Abstract:** In the field of Artificial intelligence Image processing plays a vital role in Decision making .Now a day's Mobile robots work as a Network sharing Centralized Data base.All Image inputs are compared against this database and decision is made.In some cases the Centralized database is in other side of the globe and Mobile robots compare Input image through satellite link this sometime results in delays in decision making which may result in castrophe.This Research paper is about how to make image processing in mobile robots less time consuming and fast decision making.This research paper compares search techniques employed currently and optimum search method which we are going to state.Now a days Mobile robots are extensively used in environments which are dangerous to human beings.In this dangerous situations quick Decision making makes the difference between Hit and Miss this can also results in Day to day tasks performed by Mobile robots Successful or Failure.

## Introduction

**Due to Application of Mobile robots in Diversified Fields their control and Decision making plays critical role to complete any assigned task .Now a day's Mobile Robots are applied in various fields like Military, Medical, Aeronautics etc and Decisions are taken after processing input which can in the form of image entering through its Cameras or Audio input, and through many types of Sensors. This input is compared against already existing knowledge base this knowledge base some time is centralized and all mobile robots connect it through wire or wireless in either case Input need to processed very quickly so that Mobile robot can Reach and execute the Decisions made Quickly and accurately.**

### Related Work

1) Da Vinci Surgical Robots which are surgical robot which simulates surgeon hands. These Robots are used in Medical Field for doing surgeries and they primarily depend on input compared against existing knowledge base to take decisions for what it needs to do. [1]
2) Department of Defense (DoD) Robots.[2]

The robots are used in defusing explosives after they compare the input they get from their sensors and the slight delay in decision making can result in catastrophe For the robot itself and its team [2]. These robots are used in Defense Field.

### Problem Description

The above mentioned robots Compare the input and take decision after visiting Remote database every time. This process may be ok for applications which are not that critical but for application where Decision making is critical this approach may not be best suited[1].
a. Mass communication between station and robots which overwhelm transmission media with messages causing message loss.[1]
b. Inefficient on making critical time sensitive decisions because of time consumption [1]

The problem is we have performance degradation using previous methodologies. The conventional robots use method 1 in which every input is compared with existing data base which results in more search time and less efficient in decision making capability [2].

### Suggested Solution

We will target reducing the mass communication and traffic between robots and knowledge base using cache independent memory at individual robot. Our method has many benefits such as:

1. increased decision making capability
2. Less comparisons needed to make decisions
3. Less time in processing an image.

In order to enhance the performance, we will set hit counters and will update the same in cache. The cache will have maximum to minimum hits ordered by maximum hits on the top.

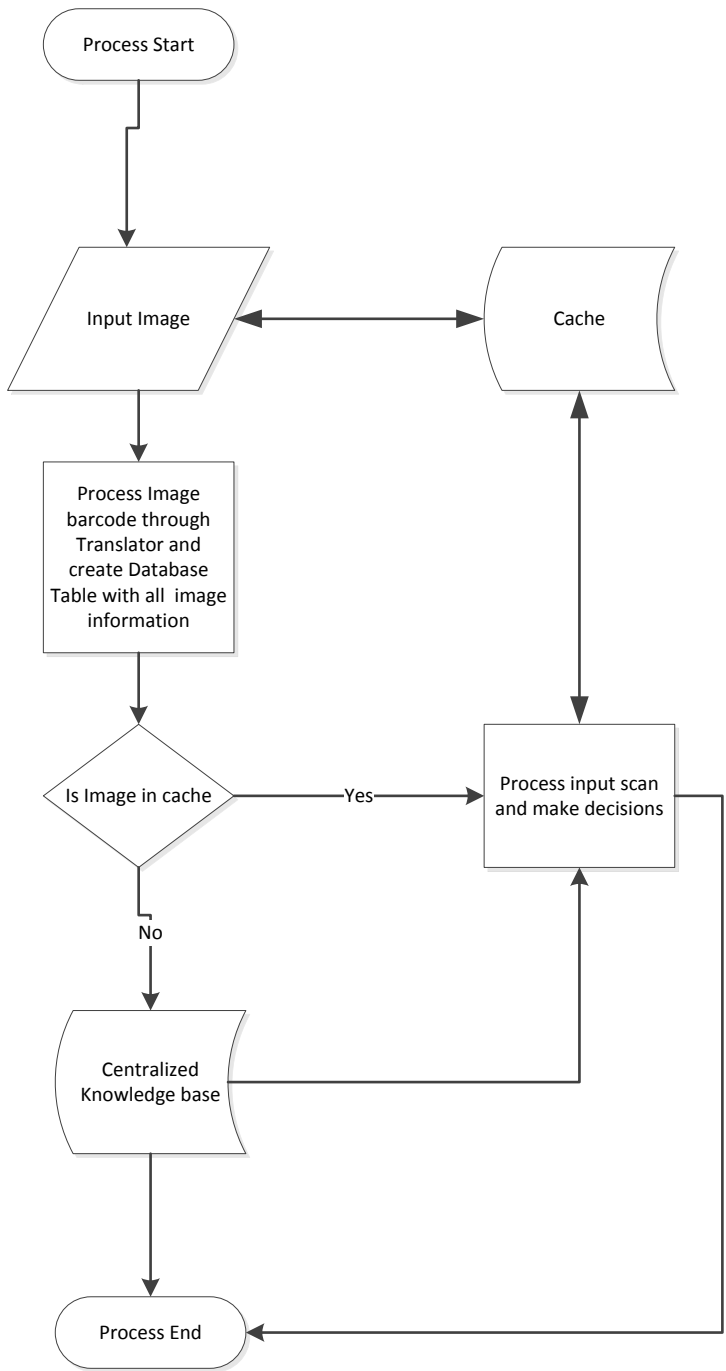

Fig. 1 Flow chart for Bhaskara & Fathima-Ibrahim Method

Case Study

In our environment, we have robots (including cache memory), satellite, station (including database).

Performance Measurements

| Input 30 Million Barcodes Scanned Images | Robots Without Cache | Recommended Bhaskara & Fathima-Ibrahim Method. |
|---|---|---|
| Decision Making Capability | 2 Minutes from time Image(Barcode) is First scanned. | 1.3 minutes from the time Image(Barcode) is First Scanned. |
| Average Time for Image Processing | 18 Minutes for Processing 30 Million Images | 15 Minutes for Processing 30 million Images |
| Message Loss or Resource Allocation | Around 1 time Resource lock or Object Lock for every 1.5 Million Images. | Around 1 time Resource lock or Object Lock for every 2.7 Million Images. |
| Number of Comparisons | 35 Millions this Includes Duplicate Images | Around 27 Million Comparisons. |

Assumptions and calculations

1) Input 30 millions Images (Barcodes) are converted into a DB2/400 file. This is done by extracting the barcode image contents Through Translator this Table is created with columns like barcode, shipper number, service type, destination terminal, delivery exceptions etc.
2) RPG language was used connect to db2/400 Database Located in Canada and is used to simulate 2 Methods.
3) Cache is simulated by array and data structure which is used to store more frequently used images hence reducing read and fetch cycles instead of going to Database located in Canada.
4) Bhaskara-Ibrahim Method is simulated by a program written in ILERPG program on AS/400 system.
5) If scanned images did not make it to our location by calculating download time by 20 minutes then we generate an alert message. This will measure time between Input obtained and processed download. Same formula is used to measure 2 Methods.

Files Declaration

```
FIMGINPUT   IF   E        K DISK
 F
RENAME(SCANNED:IMPINP30)

 FProcessfile   IF A E       K DISK
prefix(scn)

 FCompfile    UF   E       K DISK

 FRouting     O   E         DISK
```

Data Structure Declaration

```
D           DS
D  Barcode         1   14 0
D  Location        1    4 0
D  Destination     7   14 0

Array Declaration

* Initial Hits Array

D Cache      S       3 DIM(99999999)

* Final Updated Hits Array

D HoldingAreaS      3 DIM(999999999)

 * Final Updated Hits Array
```

Image scans Comparing code

```
Do while %notEof(IMGPF030)
if        IMGSCAN = 'Y'
          or IMGFOUND = 'Y'
          or NOTINARRAY = 'Y'
          or ROUNTINGS = 'Y'
MOVE      'Y'            IMGFLG
EXIST     LOOKUP         CACHE(I)
           Eval Found= 'Y'
Z-ADD     SCANDTE        IMGDTE
endif
MOVE      SCANTYP        SNDTYP
MOVE      SCANDATE       SCNDATE
MOVE      SCANTIME       SCNTIME
monitor
WRITE     IMGPF30
WRITE     INITIALHIT(I)
on-error  01021
Endif
Enddo.
```

Future Work

In the next step, we are planning is to design multilayer robots interacting with each other, updating each other cache and sharing the information.

## Acknowledgements

I thank my guide Dr. Bhaskara Reddy, Associate Professor in the department of computer Science and technology at Sri Krishnadevaraya University. I thank my colleague Mohamed Farag, Post graduate student in Maharishi University of Management in Iowa. I thank my parents, my wife and my son for their great support.